# A Stage-wise Decision Framework for Transportation Network Resilience Planning

Weili Zhang[a,e], Naiyu Wang[b,e*], Charles Nicholson[c,e] and Mohammad Hadikhan Tehrani [d,e]

[a] Graduate Research Assistant, Department of Industrial and Systems Engineering, University of Oklahoma, USA {weili.zhang-1@ou.edu}

[b] Assistant Professor, School of Civil Engineering and Environmental Science, University of Oklahoma, USA {naiyu.wang@ou.edu};

[c] Assistant Professor, Department of Industrial and Systems Engineering, University of Oklahoma, USA { cnicholson@ou.edu}

[d] Graduate Research Assistant, School of Civil Engineering and Environmental Science, University of Oklahoma, USA { m.h.tehrani@ou.edu }

[e] NIST-funded Center of Excellence on Riske-based Community Resilience Planning

*Corresponding author

**Abstract**: This study introduces a comprehensive stage-wise decision framework to support resilience planning for roadway networks regarding pre-disaster mitigation (Stage I), post-disaster emergency response (Stage II) and long-term recovery (Stage III). Three decision metrics are first defined, each based on a derivation of the number of independent pathways (IPW) within a roadway system, to measure the performance of a network in term of its robustness, redundancy, and recoverability, respectively. Using the three IPW-based decision metrics, a stage-wise decision process is then formulated as a stochastic multi-objective optimization problem, which includes a project ranking mechanism to identify pre-disaster network retrofit projects in Phase I, a prioritization approach for temporary repairs to facilitate immediate post-disaster emergency responses in Phase II, and a methodology for scheduling network-wide repairs during the long-term recovery of the roadway system in Phase III. Finally, this stage-wise decision framework is applied to the roadway network of Shelby County, TN, USA subjected to seismic hazards, to illustrate its implementation in supporting community network resilience planning.

# 1. Introduction and Background

The economy and social well-being of a community heavily rely on the availability and functionality and of its critical infrastructure systems, including power, water, gas, and transportation (PCCIP, 1997). Roadway networks are a fundamental component of transportation systems and, in the event of an extreme hazard, play a critical role during and after the event. For example, prior to the landfall of a hurricane, roadways are crucial for population evacuation; during a flood event, available roadways may provide key means of rescue; and during the longer-term recovery, the accessibility of schools, businesses, centers of government and commerce, etc. are significant elements of the economic recovery and social well-being of a community. However, the components of a roadway network, i.e. roads and bridges, are directly vulnerable to extreme hazard events; for instance, the Wenchuan earthquake in 2008 damaged 1,657 bridges in China (Zhuang et al., 2009); Hurricane Irene in 2011 affected over 500 miles of highways, 2,000 miles of roadways, and 300 hundred bridges in Vermont (Lunderville, 2011). Such physical damage can lead to extensive and expensive functionality losses to the impacted community. For example, Hurricane Sandy in 2012 caused $7.5 billion damage in direct damage to the New York transportation infrastructure (WABC-TV/DT, 2012), which have not included the indirect loss of lives, commerce, or other losses associated with the inability to effectively access emergency facilities or services, to route repair crews, or provide access to places of business. Enhancing transportation network resilience to these natural hazards has become a national imperative (Newman et al., 2011).

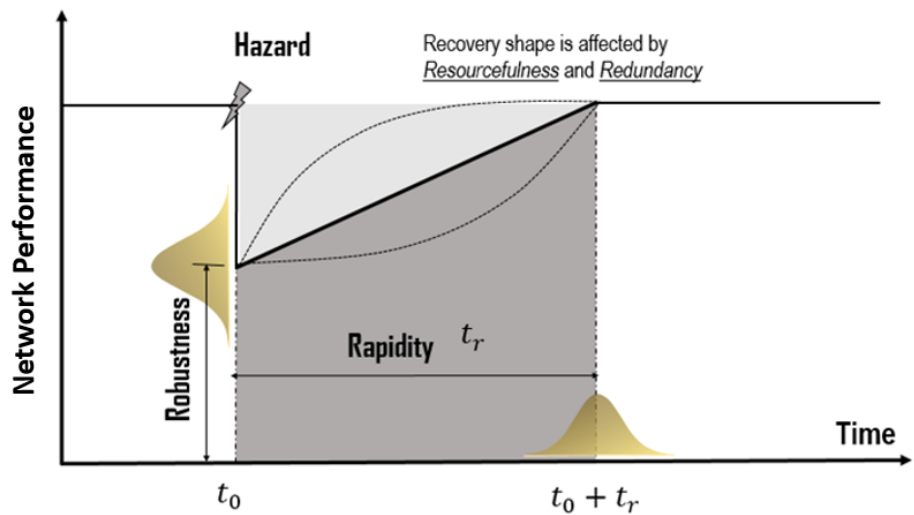

**Figure 1: Illustration of resilience concept [Source: Zhang et al. (2017)]**

Resilience - defined as the ability of social units (e.g., organizations, communities) to mitigate hazards, contain the effects of disasters when they occur, and carry out recovery activities in ways that minimize social disruption and mitigate the effects of future hazards - can be measured, as shown in Figure 1, by four major attributes: robustness (ability to withstand extreme events), rapidity (speed of recovering from impacts of extreme events), redundancy (alternative or backup components within the system), and resourcefulness (availability of emergency resources to respond to disasters) (Bruneau et al., 2003). A notable growing body of literature has been published to evaluate the performance of transportation systems from the perspective of network resilience to natural hazards (Chang and Nojima, 2001; Murray-Tuite, 2006; Ta et al., 2009; Cox et al., 2011; Frangopol and Bocchini, 2011; Ip and Wang, 2011; Cetinkaya et al., 2015; Gillen & Hasheminia 2016; Zhang and Wang, 2016; Zhang et al., 2017a, 2017b, 2018; Zhang and Nicholson, 2016a). Research studies have recently extended to decision formulations regarding risk mitigation, response and recovery of roadway networks at a community or regional scale. Bell et al. (2008) used travel time reliability as the network performance indicator to study the transportation vulnerability under terrorist attacks, in which decisions regarding the path used in the road network were made to minimize the maximum of expected loss from a predefined disruption. Bell et al. (2017) proposed capacity weighted spectral partitioning to identify the bottleneck of transportation network, which does not require information of origin-destination matrix and path assignment. Chang et al. (2010) developed a decision method for post-disaster evacuation using a proposed transportation simulation model which was capable of capturing the change of traffic pattern following seismic events. Miller-Hooks et al. (2011) proposed a two-stage stochastic programming model to maximize transportation network resilience, defined as the expected traffic demand supplied following a disastrous event, by enhancing pre-disaster preparedness and optimizing post-disaster scheduling. This model was later extended in Faturechi and Miller-Hooks (2014) to a three-stage, bi-level stochastic mathematical program, in which the upper-level is a sequence of actions for both pre-disaster mitigation and post-disaster restoration, and the lower-level reconstructs the routine for affected users using partial user equilibrium model. Sheu (2014) presented a survivor perception–attitude–resilience conceptual model with the focus on maximizing the survivor resilience after disaster. Sheu and Pan (2014) integrated three sub-networks (shelter network, medical network, and distribution network) to build a seamless centralized emergency supply network in response to natural disasters. Zhang et al. (2017) optimized roadway network recovery from an earthquake event by incorporating network topology, redundancy, traffic flow, damage states and available resources into a stochastic decision processes, resulting in an optimal schedule for sequencing restoration interventions for all damaged bridges, which lead to the fastest (in terms of time) and most efficient (in terms of indirect loss) network recovery process.

Reviewing these studies has revealed the following: i) most research studies only investigated the decision formulation or process *at one point in time* with respect to the occurrence of a hazard, and ii) different studies selected *different network performance metrics* as the basis for their decisions at hand. It has become apparent that the time at which a decision is made (with respect to the hazard occurrence) determines the situation (or constraints) and purpose (or objectives) of the decision process. It is therefore necessary for a decision framework to reflect these dynamic perspectives across the different stages of resilience planning through carefully defined performance metrics, interventions, objectives and constraints. There is an increasing need for a consistent methodology to support stage-wise resilience planning of roadway networks as more and more communities are beginning to incorporate resilience as a concept in their day-to-day risk management practices.

## 2. Stage-wise Decision Framework

The resilience concept, as illustrated in Figure 1, clearly has a time dimension. On one hand, risk mitigation and disaster management interventions at different points in time with respect to the occurrence of hazard are very different, necessitating different decision support (NIST, 2016). On the other hand, these decisions, although made at different planning stages, should serve the same overarching goal of enhancing network resilience in a systematic fashion. We propose a stage-wise decision framework, as illustrated in Figure 2, to support resilience planning of the transportation network in three phases: Stage I – pre-disaster mitigation; Stage II – post-disaster emergency response; and Stage III – long-term recovery; i.e.:

- Stage I refers to the mitigation planning stage prior to a disaster. An often-used mitigation strategy for roadway networks is to retrofit network components (i.e. bridge and roads), with an objective of enhancing network *robustness* (cf. Figure 1) and subsequently reducing damage and functionality loss of the network when a hazard event occurs. This pre-event mitigation decision problem concerns how many and which network components should be selected for retrofit subject to the available finances and other resource limitations. We propose, for Stage I, a prioritization mechanism for retrofit interventions that maximizes the network robustness under budget constraints.
- Stage II refers to the chaotic emergency response stage during or shortly after the occurrence of a hazard event. The primary functionality of a roadway network in Stage II is to connect affected population to critical facilities, e.g. hospitals, fire stations, police stations, etc., and to provide routes enabling repair crews to access damaged components of utility (e.g. power and water) networks. This functionality requirement in Stage II is often resumed through rapid repairing or establishing temporary pathways under time pressure. We therefore present a decision method to identify the damaged network components for rapid repair that can resume, within the shortest time frame, the *connectivity* between the O-D pairs that are critical for life rescue.

- Stage III refers to the long-term recovery phase, in which the impaired network is gradually restored to its pre-disaster functionality level through repair interventions. The recovery time of the network is the time it takes to repair or reconstruct all the damaged roads and bridges, which is governed by the availability of construction resources (i.e. funding, contractors, construction materials, etc.) and how they are distributed spatially along the time steps of the recovery phase. For Stage III, we propose a scheduling approach to sequence the repair interventions constrained by the characteristic of resource availability in order to resume network functionality *efficiency* with the most desirable recovery trajectory and the shortest recovery time.

The proposed stage-wise decision framework includes two major components: 1) a stage-wise network metric (SWM) system, that is formulated to be compatible with the unique objectives of the specific decisions in each of the planning stages; and 2) a stage-wise formulation of the decisions (SWD), using the SWM system as the network performance indicators in different planning stages as the basis for decision-making. The formulations of SWM and SWD are discussed in Sections 3 and 4, respectively.

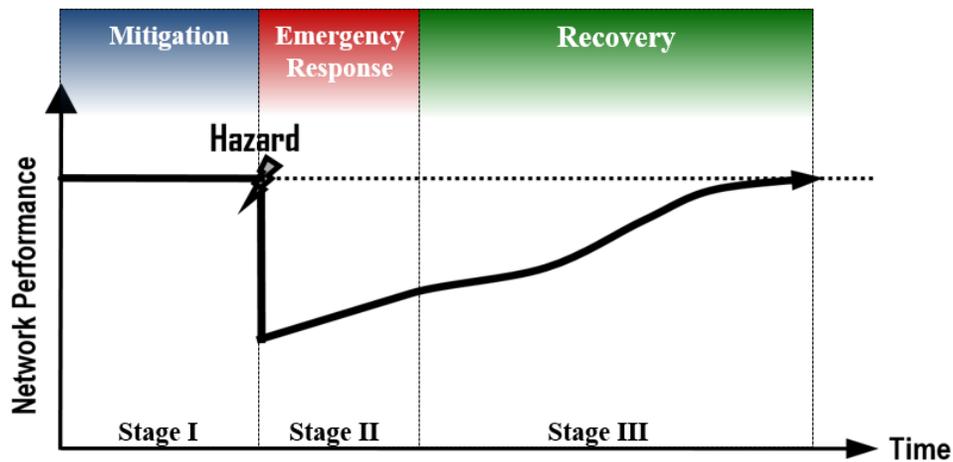

|  |  | Stage I: Mitigation | Hazard | Stage II: Emergency Response | Stage III: Recovery |
|---|---|---|---|---|---|
| **Resilience intervention** | | Retrofit bridges and/or roadway segments | | Repair or create temporary paths | Repair and rebuild the damaged network |

| Objectives | Improve the **robustness** of the network to reduce damage when hazard occurs | Ensure **connectivity** between impacted population and critical facilities to reduce fatalities and injuries | Resume full functionality and **efficiency** of the network |
|---|---|---|---|
| **Constrains** | Financial budget | Time pressure | Construction resources |
| **Network Performance Metric** | Reliability-weighted independent pathway, RIPW | Emergency facilities-weighted independent pathway, EIPW | Traffic-weighted independent pathway, TIPW |

**Figure 2: Illustration of the stage-wise decision framework**

## 3. Stage-wise Network Performance Metric (SWM) System

Different objectives of different planning phases require different network performance metrics, and different choices of metrics used in a decision formulation will ultimately result in different decisions. Cutter et al. (2008) emphasized that the most important criteria for metric selection is validity, which speaks to the question of whether the metric is representative of the resilience dimension or the decision of interest. General-purpose network performance metrics, either topology-based or functionality-based, usually cannot appropriately reflect the specific objectives of the decisions in each of the resilience planning stages. In particular, topology-based metrics (e.g. connectivity, density, etc.) alone do not reflect network functionality requirements which often are the focuses of resilience-based decisions; on the other hand, functionality-based metrics (e.g. travel cost, maximum traffic capacity) ordinarily require "too much" information (e.g. pre- and post-event network supply-demand at different spatial and temporal resolution), which may not be readily available and, more importantly, may not be directly relevant or representative of the specific objectives of the decisions to be made. That is, a metric that incorporates "too much" information could lose its sensitivity as a performance indicator to a specific dimension of network performance that is directly relevant to the current decision. To best support the stage-wise resilience planning of roadway networks, we first develop a stage-wise network performance metric (SWM) system that includes three metrics all based on the concept of independent pathways (IPW): reliability-weighted IPW (RIPW), emergency facility-weighted IPW (EIPW), and daily traffic-weighted IPW (TIPW), to support decisions in Stage I, II and III, respectively.

Edge-independent pathways, i.e. IPW, referred to in the SWM system, is an important concept in network science. The set of all IPW's in a network is defined as all paths between a two distinct nodes that do not share any common edges (e.g., road segments) (Newman, 2010). IPW provides a measure of network redundancy, which is an essential topological feature of roadway networks that strongly and positively correlates to network resiliency as observed in the 1989 Loma Prieta earthquake in California (Ardekani, 1992). Recognizing that redundancy in transportation network enhances resiliency, we employ IPW[1] is as a common substrate for the SWM system; and depending on the specific objective of the decisions in each stage, the network functionality-specific and component-specific attributes are weighted into the formulation of the IPW-based SWM system, as presented next.

We describe a transportation network on a directed graph $G = (V, A)$ where $V = \{1, 2, \ldots, n\}$ is the set of nodes and $A = \{1, 2, \ldots, m\}$ is the set of edges that represent road segments without or with maximum of one bridge[2]. The set $V$ is partitioned into two mutually exclusive sets $E = \{1, 2, \ldots, e\}$ comprised of emergency nodes (representing critical facilities, e.g., fire stations and hospitals) and $N = \{e+1, e+2, \ldots, n\}$ of non-emergency nodes (representing major destinations, e.g., residential areas, economic hubs, and major road intersections). Let $K_{ij}$ and $P_{ij}^k$, respectively, denote the total number of IPWs and the $k^{\text{th}}$ IPW between node $i \in V$ and $j \in V$. Each IPW is a set of ordered edges connected in series. An algorithm for computing $K_{ij}$ and searching for $P_{ij}^k$ between all node pairs can be found in Zhang and Wang (2016). Let $IPW_i$ denote the average number of IPWs between node $i \in V$ and all the other $n-1$ nodes, i.e.,

$$IPW_i = \frac{1}{n-1} \sum_{j=1}^{n-1} K_{ij} \tag{1}$$

The IPW of network $G$ is defined as the overall average of all $IPW_i$ values for every node $i \in V$,

$$IPW(G) = \frac{1}{n} \sum_{i=1}^{n} IPW_i \tag{2}$$

Note that IPW is independent of traveler behavior and origin-destination (O-D) demand values and can be computed efficiently using either Dijkstra's algorithm or the Ford-Fulkerson algorithm for max flows (Ahuja et al. 1993).

## 3.1 Reliability Weighted IPW, RIPW

---

[1] As a redundancy measure in the context of resilience-based decision for road networks, IPW is preferred to all-possible-pathway (APW) for two reasons: (a) finding APW itself is NP-hard (Ausiello et al., 2012), which makes it impractical for real-world problems, while IPWs is limited and can be easily enumerated by Dijkstra's (Skiena, 1990) or the Ford-Fulkerson algorithm (Ahuja et al., 1993); and (2) since many paths could share a common edge, APW can be overly sensitive to certain local damages.

[2] An edge is a segment that can only have one bridge at most, because decision discussed in this study are made on individual bridges, each of which need a unique identity in the network topology representation. For a road with two bridges, it can be easily partitioned to two consecutive roads, each of which has one bridge.

In Stage I, the purpose of pre-disaster mitigation through component retrofit is to enhance the network robustness as discussed in Section 2, i.e. to reduce the network damage when hazard occurs. Increasing reliability (or reducing failure probability) of selected network components to minimize the possible damages to the IPWs between all O-D pairs is the objective of retrofit interventions. Accordingly, we define reliability-weighted IPW, RIPW, as the network performance metric to support the Stage I pre-event mitigation decisions. Assuming conservatively[3] that reliabilities of edges are statistically independent, the reliability of path $P_{ij}^k$, denoted as $R_{ij}^k$, is the product of the reliabilities of all edges included in the path $P_{ij}^k$

$$R_{ij}^k = \prod_{l \in P_{ij}^k} q_l \tag{3}$$

where $l$ and $q_l$ are the edge index and the corresponding reliability. The average number of reliable IPWs for node $i \in V$ is denoted as $Q_i^R$ and calculated as shown in Eq (4):

$$Q_i^R = \frac{1}{n-1} \sum_{j=1}^{n-1} \sum_{k=1}^{K_{ij}} R_{ij}^k \tag{4}$$

Let $Q^R(G)$ denote the RIPW of network $G$ which is computed in Eq (5):

$$Q^R(G) = \frac{1}{n} \sum_{i=1}^{n} Q_i^R \tag{5}$$

A critical step to compute RIPW is to quantify reliabilities of the edges (i.e., road segments and bridges), which depend on the specific hazard of interest, and often are evaluated through fragility analyses associated with damage states of interest [e.g. HAZUS-MH MR4 (2009)].

## 3.2 Emergency Node-Weighted IPW, EIPW

In Stage II, immediately after a disruptive event, emergency managers need to quickly identify and establish emergency routes to send rescue teams and relief resources to the affected population. The connectivity between critical facilities and the people who need to be rescued is of the paramount concern (Jones and Bentham, 1995; Higgs, 2004). Accordingly, we define the emergency facility-weighted IPW, EIPW, as the network performance metric for Stage II. EIPW measures the average number of independent pathways between emergency nodes and non-emergency nodes. The EIPW is highly dependent on the post-event damage condition of the network. Let $d_{ij}$ denote the damage status of edge $(i,j) \in A$, which can be measured on a 0 to 4 scale, corresponding to the damage level of none, slight, moderate, extensive and complete, respectively, as termed in HAZUS MH-2.2 (2015). Edges with the damage conditions of extensive or complete are considered unserviceable. The serviceability of edge $(i,j) \in A$ is denoted $s_{ij}$ and defined as,

$$s_{ij} = \begin{cases} 1, & d_{ij} \leq 2 \\ 0, & \text{otherwise} \end{cases} \quad (6)$$

---

[3] Although this assumption regarding may not hold exactly for natural hazards with large geographic footprints, it is well-known from systems reliability theory that this assumption is conservative, especially for bridges that are widely separated in the system.

Eq. (7) defines the serviceability of $P_{ij}^k$, denoted as $s_{ij}^k$, as the product of the serviceability of all road edges included in the path $P_{ij}^k$:

$$s_{ij}^k = \prod_{l \in P_{ij}^k} s_l \quad (7)$$

If $s_{ij}^k$ equals 1, the path is serviceable and if it equals 0, it is not. Let $Q_i^E$ denote the EIPW of node:

$$Q_i^E = \frac{1}{e} \sum_{j \in E} \sum_{k=1}^{K_{ij}} s_{ij}^k \quad (8)$$

The EIPW of a network is defined as the average EIPW of non-emergency nodes, which is computed as,

$$Q^E(G) = \frac{1}{n-e} \sum_{i \in N} Q_i^E \quad (9)$$

If $Q^E(G)$ equals 0, the transportation network cannot support any emergency actions.

### 3.3 Average Daily Traffic Weighted IPW, TIPW

In Stage III, we use the pre-event average daily traffic (ADT) weighted IPW, or TIPW, as the network performance metric fundamental to decisions. The ADT on roads and bridges are the field measurements routinely maintained by the Federal or State Department of Transportation. ADT data may be more accurate than current traffic assignment models at representing real traffic patterns since the path choice of travelers are only *partially* rational, whereas traffic assignment models commonly assume decisions to be based on distance or travel time (Yagar, 1971; Patricksson, 1994). We emphasize that we use *pre-event* ADT data as the benchmark for recovery decisions, as the goal in Stage III is to resume the full *pre-event* functionality level of the network through a timely and effective recovery process. The service level of each edge $(i,j) \in A$ is idealized as $1 - \frac{d_{ij}}{4}$. For example, if an edge is completely damaged, $d_{ij}$ is set to be 4, and the corresponding service level is 0. The service level of the path $P_{ij}^k$ is approximated as the product of the service levels of all arcs $l \in P_{ij}^k$. Let $A_l$ denote the ADT of edge $l \in P_{ij}^k$. Define $A_{ij}^k$, the ADT of $P_{ij}^k$, as the minimum ADT of all edges on that pathway, i.e.,

$$A_{ij}^k = \min \{A_l | l \in P_{ij}^k\} \tag{10}$$

The normalized ADT of the path is defined as:

$$\tilde{A}_{ij}^k = \frac{K_{ij} A_{ij}^k}{\sum_{p=1}^{K_{ij}} A_{ij}^p} \tag{11}$$

Note that for any node pair $i, j \in V$, $\sum_{k=1}^{K_{ij}} \tilde{A}_{ij}^k = K_{ij}$. Let $Q_i^A$ denote the TIPW of node $i \in V$ and

$$Q_i^T = \frac{1}{n-1} \sum_{j=1}^{n-1} \sum_{k=1}^{K_{ij}} \tilde{A}_{ij}^k \prod_{l \in P_{ij}^k} (1 - \frac{d_l}{4}) \tag{12}$$

The TIPW of a network is further defined as:

$$Q^T(G) = \frac{1}{n} \sum_{i=1}^{n} Q_i^T \tag{13}$$

For Stage III, a damaged transportation network is considered fully recovered if all the damaged roadway segments are restored, i.e., the network TIPW computed using Eq. (13) returns to its pre-disaster level (without damaged network components).

The defined SWM – comprised of RIPW, EIPW and TIPW – are tailored and sensitive to the unique objectives of the specific decisions they are serving, at the same time, they ensure a level of consistency among the decisions in different planning stages. These IPW-based metrics can be employed individually to support decisions at each planning stage, or used together as a system in the SWD formulated in Section 4.

## 4. Formulation of the Stage-wise Decision (SWD) Framework

### 4.1 Stage I: Pre-disaster Risk Mitigation

At Stage I, the decision is to make selections from a set of candidate edges to maximize the RIPW and simultaneously minimize the related investment cost. The mitigation decisions, represented as the binary decision vector $\boldsymbol{x}$, is comprised of a component $x_{ij}$ for each $(i,j) \in A$ as defined as below,

$$x_{ij} = \begin{cases} 1, \text{edge } (i,j) \text{ is selected} \\ 0, \quad\quad\quad\quad\quad \text{otherwise} \end{cases}, \forall (i,j) \in A \quad (14)$$

The first objective of the decision process is to maximize the expected RIPW:

$$\max Q^R(\boldsymbol{x}) = E_{\hat{\xi}_q} \left\{ \frac{1}{n} \sum_{i=1}^{n} \frac{1}{n-1} \sum_{j=1}^{n-1} \sum_{k=1}^{K_{ij}} \prod_{l \in P_{ij}^k} [(1-x_l) q_l(\hat{\xi}_q) + x_l q_l'] \right\} \quad (15)$$

where $q_l(\xi_q)$ is the realization of the random variables $q_l$ representing the reliabilities of edges $l \in P_{ij}^k$ prior to retrofit, and $q_l'$ is the reliability of edge after retrofit. Eq. (15) ensures that the reliability of edge $l$ is $q_l'$ if it is selected; otherwise, it does not change. Let $C(\boldsymbol{x})$ denote the total cost associated with decision $\boldsymbol{x}$. The second objective is to minimize the expectation of total retrofit costs:

$$\min C(\boldsymbol{x}) = E_{\hat{\xi}_c} \left[ \sum_{(i,j) \in A} c_{ij}(\hat{\xi}_c) x_{ij} \right] \quad (16)$$

where $c_{ij}(\hat{\xi}_c)$ is the stochastic realization of random variable retrofit cost $c_{ij}$.

This decision problem is closely related to the stochastic knapsack problem (Ross and Tsang, 1989), but further complicated by the procedure of iteratively computing a series of weighted shortest paths. The non-dominated sorting genetic algorithm II (NSGA-II), successfully applied to several similar problems (Chu and Beasley, 1998; Deb et al., 2002), is employed to identify the near-optimal Pareto frontier for this multi-objective optimization problem in which the two objectives conflict. We couple NSGA-II with Monte Carlo Simulation (MCS) and optimize the expected RIPW and total costs.

### 4.2. Stage II: Post-disaster Emergency Response

In this stage, the problem is to select a number of edges from failed edges and to schedule their restoration sequence, the objective is to minimize the total time to ensure that there exists at least one path between non-emergency nodes and emergency nodes. Let $D = \{1, 2, \dots, b\}$ denote the set of damaged edges (candidates). Let $x_l$ and $y_l$ respectively denote whether or not to select edge $l \in D$ and the time at which emergency restoration is initiated for edge $l \in D$. The network emergency recovery time associated with the selection $\boldsymbol{x}$ and schedule $\boldsymbol{y}$ is denoted as $t^E(\boldsymbol{x}, \boldsymbol{y})$. Let $s_l'$ and $p_l(\hat{\xi}_p)$ denote serviceability for each edge $l \in D$ after emergency restoration and the stochastic variables of duration of temporal action. Let $T$ denote the set of discrete points in time: $\{t_0\} \cup \{t_1, \dots, t_b\}$ which is the union of initial time $t_0$ and the intervention completion times $t_l = x_l[y_l + p_l(\hat{\xi}_p)]$ for each edge $l \in D$. Using the above notation, the emergency accessibility recovery model is presented as follows:

$$\min t^E(\boldsymbol{x}, \boldsymbol{y}) = E_{\hat{\xi}_p} \langle \max_{l \in D}\{x_l[y_l + p_l(\hat{\xi}_p)]\} - t_0 \rangle \tag{17}$$

subject to:

$$\sum_{j \in E} \sum_{k=1}^{K_{ij}} \prod_{l \in P_{ij}^k} s_l{'} \geq 1, \forall i \in N \tag{18}$$

$$s_l{'} = (1 - x_l)s_l + x_l, \forall l \in D \tag{19}$$

$$\sum_{l \in D} [t \geq y_l][t \leq y_l + p_l(\hat{\xi}_p)] \leq N_{SL}^{max}, \forall t \in T \tag{20}$$

$$y_l \leq Mx_l, \forall l \in D \tag{21}$$

$$s_l{'}, x_l \in \{0,1\}, \forall l \in D \tag{22}$$

$$y_l \geq 0, \forall l \in D \tag{23}$$

Eq. (17) defines the total emergency accessibility recovery time as the intervention completion time of the very last scheduled edge. Eq. (18) requires each non-emergency node to have at least one IPW to an emergency node and Eq. (19) ensures the edge $l \in D$ remains at its initial serviceability unless it is selected for restoration. In Eq. (20), $[P]$ is the Iverson bracket, which returns 1 if $P$ is true, and 0 otherwise. $N_{SL}^{max}$ denotes the maximum number of simultaneous emergency restoration interventions possible based on the the human and financial resources available in the community for emergency recovery during this stage. The parameter $M$ in Eq. (21) is set to a large number and is used to ensure edge $l \in D$ is scheduled for restoration only if it is selected. Note that $M$ must be a sufficiently large number (e.g., greater than any reasonable scheduling times) to not unintentionally reduce the set of feasible solutions.

The problem under investigation is closely related to the NP-hard parallel machine selection and scheduling problem (Cheng and Gen, 1997; Subramaniam et al., 2000; Cao et al., 2005). We assume bridge repair scheduling is non-preemptive, that is, once a crew has begun repair on a given bridge, they must complete their work before moving to another bridge. A genetic algorithm (GA) is employed to identify the near-optimal solutions for the emergency accessibility recovery problem.

### 4.3 Stage III: Long-term Recovery

In the final stage, the objective is to restore the transportation system to the pre-disaster condition with minimized total recovery time and maximized efficiency. This paper utilizes two criteria to evaluate the rapidity and efficiency of the long-term recovery scheduling: total recovery time (TRT) and the skewness of the recovery trajectory (SRT), which are recently introduced by Zhang et al. (2017). It must be noted that the emergency restoration in Stage II is only associated with temporary interventions – that is, any bridge selected in Stage II, has not been repaired for long-term use and must be considered in Stage III. The long-term recovery problem then is to determine an optimal schedule $x = \{x_1, x_2, \dots, x_b\}$ for the repair of all $b$ damaged edges. Let $r_l(\hat{\xi}_p)$ denote the realization of the random variable $r_l$ denoting the duration of the restoration intervention on bridge $l$. Let $t^R(x)$ and $t^S(x)$ denote the TRT and SRT associated with schedule $x$, respectively. We set $T' = \{t'_0, t'_1, \dots, t'_b\}$ as the permutation of $T$ such that $t'_0 \leq t'_1 \leq t'_2 \leq \cdots \leq t'_b$. The long-term recovery model is presented as below:

$$\min t^R(x) = E_{\hat{\xi}_p} \left\{ \max_{l \in B}[x_l + r_l(\hat{\xi}_p)] - t_0 \right\} \tag{24}$$

$$\min t^S(x) = E_{\hat{\xi}_p} \left\{ \frac{\sum_{i=1}^{u} t'_i \, Q^A_{t'_i}(t'_i - t'_{i-1})}{\sum_{i=1}^{d} Q^A_{t'_i}(t'_i - t'_{i-1})} \right\} \tag{25}$$

subject to:

$$Q^A_t = \frac{1}{n}\sum_{i=1}^{n}\frac{1}{n-1}\sum_{j=1}^{n-1}\sum_{k=1}^{K_{ij}} \tilde{A}^k_{ij} \prod_{l \in P^k_{ij}}\left(1 - \frac{d^t_l}{4}\right), \forall t \in T' \tag{26}$$

$$d^t_l = d_l[x_l + p_l(\hat{\xi}_p) > t], \forall t \in T' \tag{27}$$

$$\sum_{l \in B}[t \geq x_l][t \leq x_l + p_l(\hat{\xi}_p)] \leq N^{max}_{SL}, \forall t \in T' \tag{28}$$

$$x_l \geq 0, \forall l \in D \tag{29}$$

As discussed in Section 4.1, we use NSGA-II to search the near-optimal non-dominated solutions for the multi-objective stochastic integer-programming problem. In addition, there are other techniques can be used to solve this problem (Nicholson and Zhang, 2016; Zhang and Nicholson, 2016b, 2018).

## 5. Stage-wise Resilience Planning for Roadway Network in Shelby County, TN

The proposed SWD framework is illustrated using a skeleton highway network in the Shelby County, Tennessee, illustrated in Figure 3. The system includes 46 edges representing the highways and 34 nodes representing major road intersections and economic hubs; there are 10 hospitals in the study area, 8 of which are located in downtown. There are 24 bridges are considered in this skeleton network, and the detailed descriptions of those bridges are presented in Table 1. For simplicity, we only consider bridges as the vulnerable network components in the subsequent illustration.

The chance of a moderate earthquake occurring in the New Madrid Seismic Zone (NMSZ) in the near future is high. Scientists estimate that the probability of a magnitude 6 to 7 earthquake occurring in NMSZ within the next 50 years is higher than 90% (Hildenbrand et al. 1996). However, most civil infrastructure in the NMSZ were not seismically designed, as opposed to those in frequent earthquake regions (e.g., California, USA or Japan). We consider a scenario earthquake with magnitude Mw equal to 7.7 and the epicenter located at 35.3N and 90.3W (on the New Madrid Fault Line) as proposed in the MAE Center study (Adachi, 2007). A selected ground motion attenuation model (Atkinson and Boore, 1995) is used to estimate the peak ground acceleration at the site of the bridges.

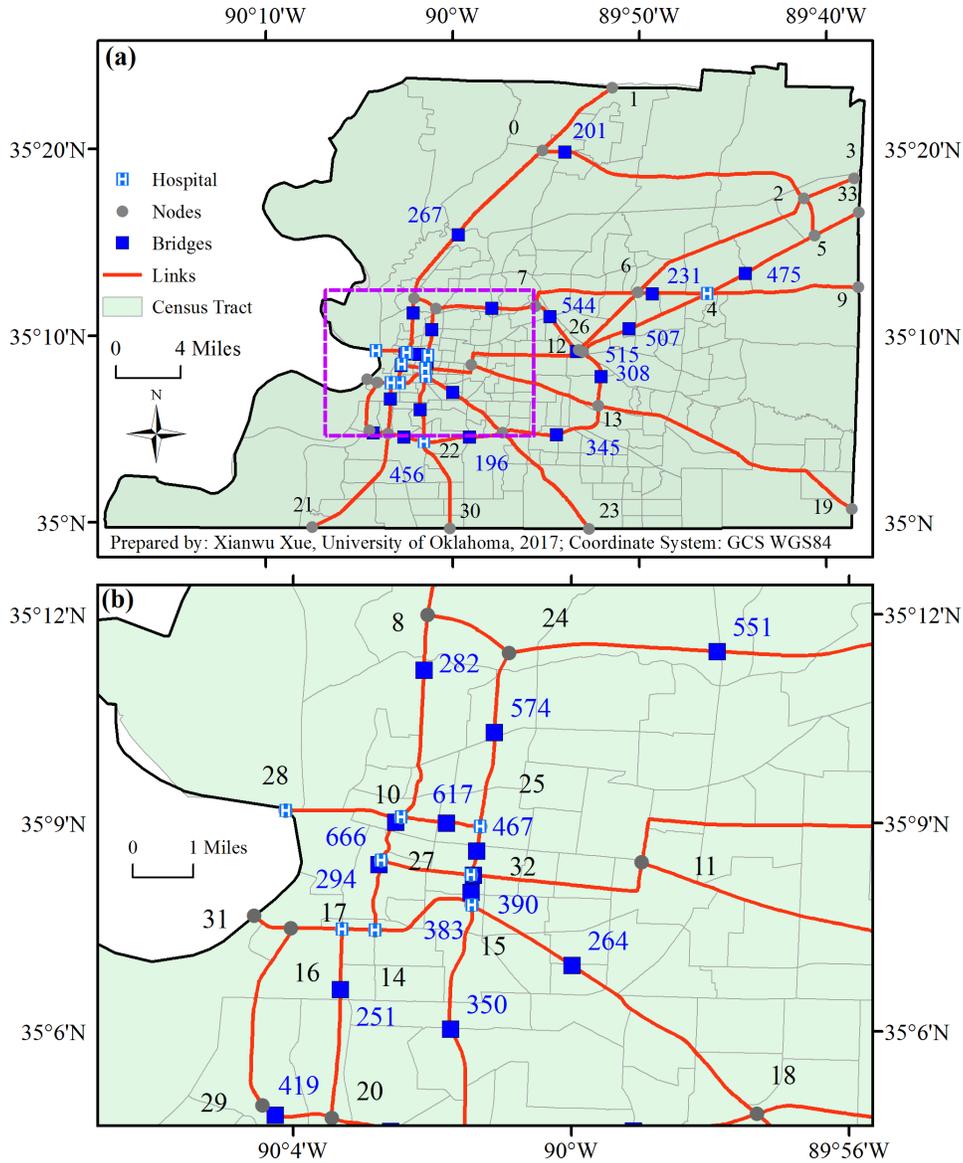

**Figure 3: (a) Road network in Shelby County, TN. (b) Detailed network topology in downtown Memphis, TN.** [The plotting area in (b) is highlighted by the dash-line box in (a)]

The random variables, including ADT ($A_{ij}$), pre-disaster retrofit costs ($c_{ij}$) (assumed to be a function of bridge geometry, material type, year built, and deck area), reliabilities ($q_{ij}$), expected damage ($d_{ij}$) and the corresponding restoration duration ($p_{ij}$), for each of the bridges under the considered earthquake scenario, are assume to be normal distributions, with mean values listed in Table 1 and an assumed coefficient of variation of 7% (the implication of this assumption, often made in probabilistic risk analysis, is that the parameters of each bridge in this network can be estimated to within ± 15% with more than 95% confidence).

Table 1: Mean values of bridge parameters

| Bridge ID | Construction Type | Deck area (m²) | Reliability | Retrofit Cost | ADT | Damage Status | Duration of Restoration Intervention (Days) |
|---|---|---|---|---|---|---|---|
| 196 | MSSS_Concrete | 19,248 | 0.774 | $30,796 | 21,500 | slight | 2 |
| 201 | MSSS_Concrete | 21,792 | 0.786 | $49,031 | 79,910 | slight | 4 |
| 231 | MSSS_Concrete | 21,648 | 0.781 | $48,708 | 113,830 | slight | 4 |
| 251 | MSC_Concrete | 24,000 | 0.740 | $28,799 | 94,920 | slight | 4 |
| 264 | MSC_Concrete | 41,424 | 0.793 | $49,708 | 94,130 | none | 0 |
| 267 | MSSS_Steel | 3,792 | 0.781 | $6,067 | 99,190 | none | 0 |
| 282 | SS_Concrete | 14,640 | 0.962 | $58,561 | 121,570 | none | 0 |
| 294 | MSC_Concrete | 35,280 | 0.887 | $141,115 | 13,260 | extensive | 82 |
| 308 | MSC_Concrete | 36,000 | 0.712 | $143,994 | 28,220 | moderate | 24 |
| 345 | MSC_Steel | 76,224 | 0.408 | $121,956 | 84,090 | extensive | 108 |
| 350 | MSSS_Steel | 74,928 | 0.638 | $89,911 | 13,770 | slight | 8 |
| 383 | MSC_Concrete | 24,432 | 0.716 | $54,971 | 13,490 | moderate | 21 |
| 390 | MSC_Steel | 48,864 | 0.347 | $78,180 | 2,040 | complete | 287 |
| 419 | MSC_Concrete | 19,920 | 0.745 | $79,675 | 99,000 | complete | 214 |
| 456 | MSC_Concrete | 31,296 | 0.841 | $125,180 | 39,110 | moderate | 22 |
| 467 | MSSS_Concrete | 36,864 | 0.900 | $44,236 | 6,450 | slight | 6 |
| 475 | MSSS_Concrete | 6,072 | 0.779 | $24,289 | 14,580 | extensive | 67 |
| 507 | MSSS_Steel | 61,152 | 0.519 | $97,841 | 11,320 | extensive | 95 |
| 515 | MSC_Steel | 42,288 | 0.280 | $50,744 | 16,890 | complete | 267 |
| 544 | MSSS_Concrete | 19,536 | 0.835 | $23,443 | 7,770 | none | 0 |
| 551 | MSC_Steel | 55,008 | 0.336 | $123,765 | 14,020 | extensive | 83 |
| 574 | MSSS_Concrete | 8,640 | 0.905 | $19,439 | 31,660 | moderate | 14 |
| 617 | MSSS_Concrete | 16,248 | 0.855 | $25,996 | 16,320 | slight | 2 |
| 666 | MSC_Steel | 76,512 | 0.482 | $306,042 | 49,190 | none | 0 |

*MSSS – multi span simply supported; MSC- multi span continuous; SS-single span

**Stage I: Mitigation Strategy**

When all the bridges in the network are subjected to standard traffic flow (the reliabilities of bridges under service loads are assumed to be 0.99), the RIPW is 1.697, meaning on average there are 1.697 reliable independent pathways (IPW) between all O-D pairs in the network under normal operational condition. When the prescribed scenario earthquake is considered, the mean network RIPW drops to 0.635, indicating many O-D pairs become disconnected following the earthquake event. Figure 4 shows the Pareto front of the optimal retrofit strategies mapped in the objective space with each triangular marker representing a non-dominated solution with respect to a specific combination of financial investment and corresponding RIPW; these non-dominated solution cannot be improved with respect to one objective function without diminishing the other.

The specifics of the four strategies on the Pareto front, highlighted in Figure 4, are summarized in Table 2. For example, Strategy I in Figure 4 is located toward the bottom left of the objective space and emphasizes cost savings over RIPW. This strategy identified three bridges (bridges 196, 390, and 515) for retrofit which are located near the center of the roadway system and likely are included in IPWs for several O-D pairs. Furthermore, bridges 390 and 515 (cf. Table 1) are among the bridges with lowest reliability but their retrofit costs are relatively moderate. Strategy II moves toward spending more money to improve RIPW and indicates that with a budget of $447K, 9 bridges can be retrofitted, leading to a RIPW of 1.00. The significance of a RIPW equal to or greater than 1 is that on average at least one reliable pathway exists between all O-D pairs. The RIPW increases for solutions along the Pareto front as we move towards the upper right of the objective space. As costs increase, more bridges can be selected for retrofit. However, the tradeoff between RIPW and retrofit cost is not simply one of adding more bridges. For instance, bridge 196 is selected in Strategy I, but de-selected in Strategy II and III because neighboring bridges (bridges 345 and 350) are more cost-effective for improving the overall network RIPW; this Bridge 196 is re-selected again in Strategy IV when the retrofit budget become sufficiently high. Additionally, some bridges are *always* selected in the optimal retrofit solutions, e.g., bridges 390 and 515, often because their reliability is much lower than average, or they are shared by IPWs between multiple O-D pairs.

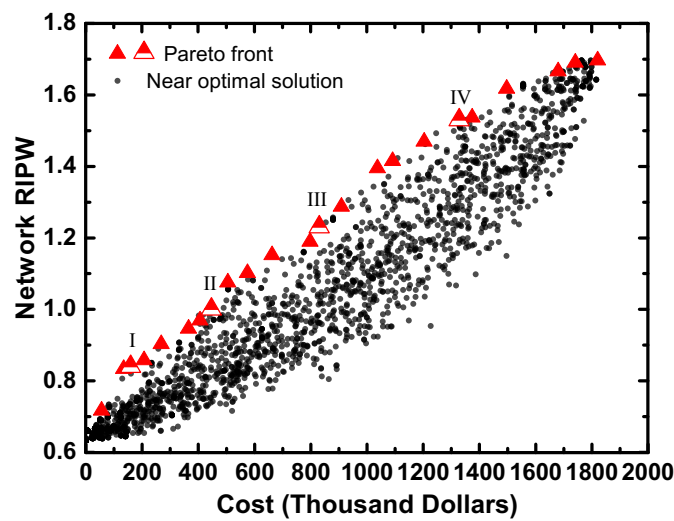

**Figure 4: Pareto front: RIPW and Cost**

Table 2: Details of four optimal solutions in Pareto front

| Strategy | Expected cost ($) | Expected RIPW | Number of bridges selected for retrofit | Bridges selected |
|---|---|---|---|---|
| I | 160K | 0.82 | 3 | 196, 390, 515 |
| II | 447K | 1.00 | 9 | 267, 350, 383, 390, 475, 507, 515, 574, 617 |
| III | 831K | 1.23 | 12 | 267, 282, 308, 345, 383, 390, 467, 475, 507, 515, 551, 617 |
| IV | 1,329K | 1.46 | 20 | 196, 201, 231, 251, 264, 267, 282, 308, 345, 350, 383, 390, 419, 456, 467, 475, 507, 515, 544, 551 |

**Stage II: Emergency Response**

The actual damage condition of the network immediately following a hazard event should serve as the initial state for the Stage II planning. For illustration, we use the mean damage states tabulated in Table 1 as the starting point. As shown in Figure 5, following the scenario earthquake, among the 24 bridges, 5 bridges sustained negligible damage, 7 bridges are slightly damaged, 4 bridges have moderate damage, 5 bridges suffered extensive damage, and 3 bridges fail completely. Prior to the scenario earthquake the EIPW is 1.89, however, immediately after it falls to 0.55. In particular, nodes 11, 13, and 19 (highlighted in Figure 5), do not have access to any hospital in the Shelby area. We assume that the time to establish a temporary path is on average 20 hours and 35 hours for extensively and completely damaged bridges, respectively, with COV of 10%. Furthermore, we assume a maximum of two repair teams (i.e. $N_{SL}^{max} = 2$) are available for emergency rescue immediately following the disaster. Our decision space is to schedule the rapid repair for the 8 severely damaged bridges in order to restore the EIPW between all non-emergency nodes and hospitals from 0.55 to a value that is greater than or equal to 1.0 as quickly as possible.

**Error! Reference source not found.** reveals the best emergency recovery time defined by Eq. (17) at each iteration of the genetic algorithm. The GA converges at the 16th iteration with an emergency recovery time of 20 hours. The optimal solution identifies that the bridge 345 is the most critical bridge to bring the EIPW of nodes 11, 13, and 19 up to 1.0 within the shortest time. These three nodes can all use bridge 345 (after emergency restoration) and 264 (with only negligible damage) to access hospitals. While bridge 390 is an alternative candidate (to bridge 345), it is completely damaged and will take longer for emergency restoration. The illustrated approach provides useful information for emergency managers to make informed decisions for timely response and relief efforts.

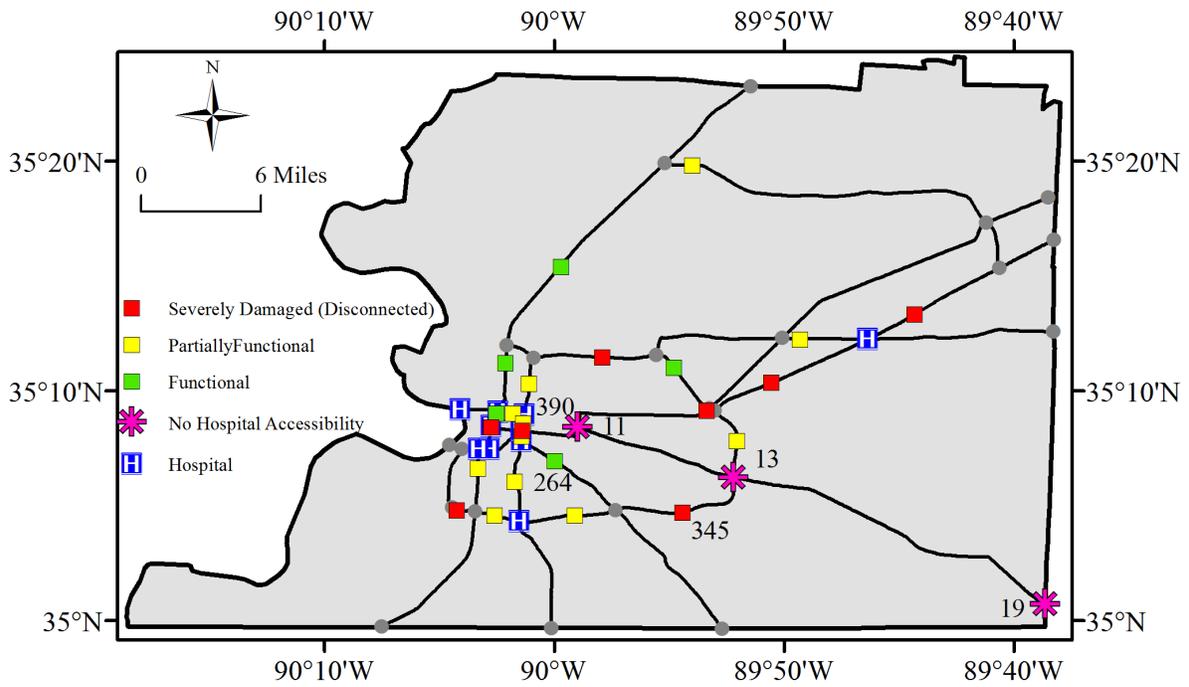

**Figure 5: Damage condition of the roadway network**

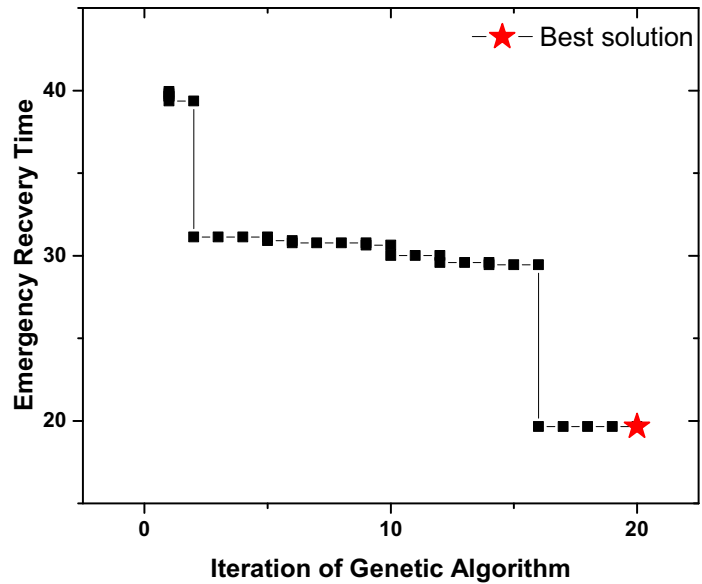

**Figure 6**: **Emergency recovery time over GA iterations**

**Stage III: Long-term Recovery Scheduling**

The network performance in Stage III is evaluated using TIPW as the performance metric defined by Eq. (13). The TIPW equals 0.578 following the earthquake event, while the pre-hazard TIPW is 1.752 under normal traffic conditions. The decision problem is to determine the restoration schedule for all the 19 damaged bridges (including slight, moderate, extensive, and complete damage states) to minimize the TRT and SRT. We assume limited resources are available and only a maximum of 5 bridges can be repaired simultaneously.

The optimal recovery trajectories are depicted in Figure 7. The optimal solution is clearly superior to the suboptimal solution shown, even though they both have the same total recovery time. For example, the optimal solution takes 88 days to improve the TIPW to 1.0, while it takes 178 days for the suboptimal solution to achieve the same results. This confirms that TRT, if it is used as the sole objective, is not sufficient to ensure the recovery schedule with the "best" trajectory. The restoration schedule corresponding to the optimal solution is detailed in Figure 8. The complete network recovery occurs in 300 days given that the maximum of 5 bridges can be repaired simultaneously. The optimal strategy selects in the early phase 8 slight or moderately damaged bridges associated with large traffic flow (e.g., the ADT of bridge 231 is 113K). The restoration of these 8 bridges improves the network TIPW by 34% in 34 days as displayed in Figure 7. Bridge 390 (which takes 287 days to restore) and the other two completely damaged bridges (515 and 419) are scheduled early along the time horizon to ensure total recovery time is minimized. These results demonstrate that the optimal scheduling of bridge restoration can greatly improve the efficiency of the transportation network recovery. Although reasonable strategies can be developed by empirical analysis, the advantage of the proposed model would become more apparent when dealing with large, extensively-damaged networks where the decision variables, possible alternative strategies and constraints create a complex decision problem where intuition may fail.

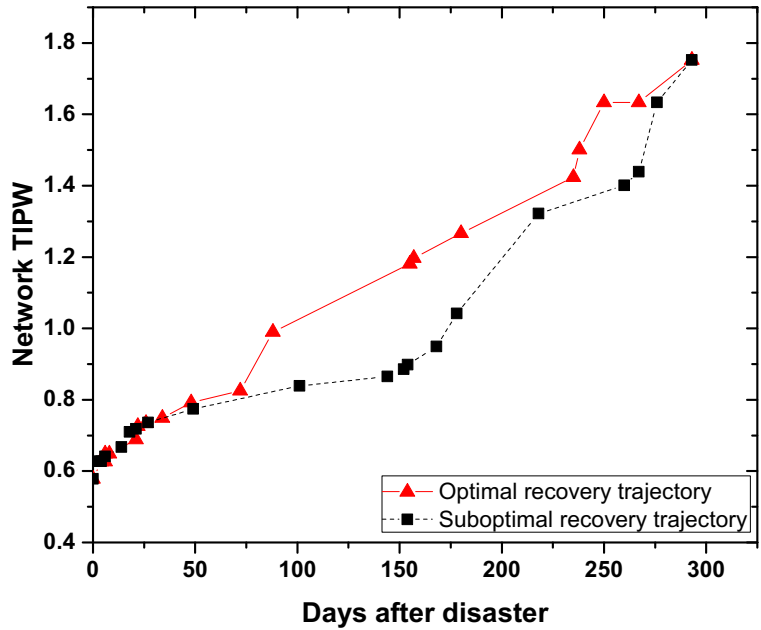

**Figure 7: Recovery trajectories with different restoration schedules**

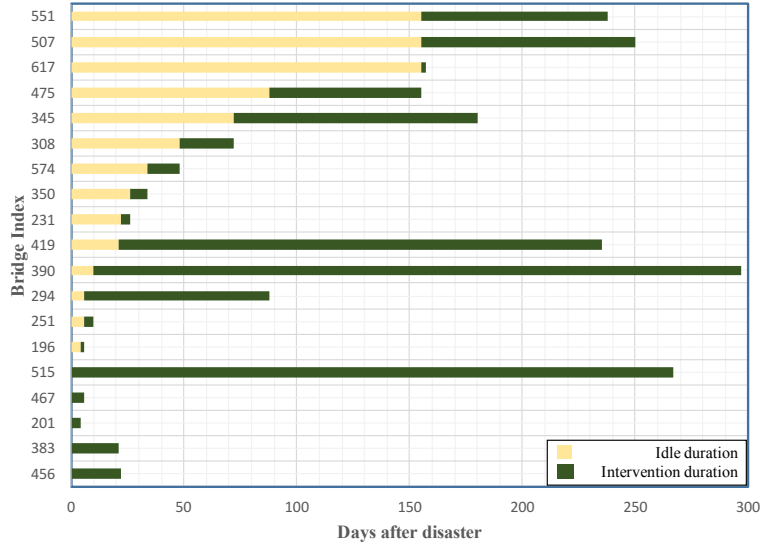

**Figure 8: The optimal scheduling for long-term recovery of the network**

## 6. Conclusions

As more and more communities are incorporating resilience as key concept in their day-to-day practices, public decision makers and other stakeholders are likely to be required to take a system perspective concerning physical infrastructure risk management. To that end, this paper formulates a comprehensive stage-wise decision framework to support the resilience planning activities for roadway network in the three main stages: pre-disaster mitigation (Stage I), post-disaster emergency response (Stage II), and long-term recovery (Stage III). The decision framework includes an IPW-based SWM system and a set of stochastic SWD formulations that are designed to meet the specific objectives and decision purposes of each of the planning phases. Specifically:

- Three quantitative resilience-based, decision-specific network performance metrics: RIPW, EIPW and TIPW are proposed. RIPW is formulated to measure the network robustness to extreme natural hazards. EIPW is designed to quantify the critical facility accessibility in damaged roadway networks. TIPW represents the efficacy of the network in carrying traffic flow. These specifically defined SWMs are used in formulating the specific objectives of the decisions in each of the planning phases.

- Three stochastic optimization models are formulated to support decisions in each planning stage. In Stage I, we develop a multi-objective stochastic binary integer programming problem to select bridges for retrofit with minimized total cost and maximized RIPW. In Stage II, we employ a stochastic integer programming problem to select and schedule emergency restoration (temporary path) for damaged bridges constrained by limited resources to ensure all residents have access to hospitals within the shortest time period following extreme events. In Stage III, a multi-objective stochastic integer programming problem is built to optimize the schedule of network recovery interventions by minimizing the total recovery time (TRT) and maximizing the efficiency (SRT) during the recovery period under resource constraints. All these decision problems are NP-hard; the NSGA-II and GA are designed to identify near optimal solutions efficiently and are scalable to handle large and complex networks.

## Acknowledgement

The research reported herein was supported, in part, by the Center for Risk-Based Community Resilience Planning, funded by the National Institute of Standards and Technology (NIST) under Cooperative Agreement No.70NANB15H044. This support is gratefully acknowledged. Thanks are extended to Dr. Xianwu Xue for generating the GIS plots of the Shelby roadway networks.